\documentclass[12pt]{article}
\usepackage{graphicx,rotating}
\usepackage{cite}
\newcommand{\HRule}{\rule{0.4\linewidth}{0.3mm}}
\newcommand{\mlab}[1]%
    {\mbox{}\marginpar{\raggedright\hspace{0pt}\footnotesize #1}}

\newcommand{\jpsi}{J/$\psi$}
\newcommand{\psip}{$\psi^\prime$}

\newcommand{\dd}{{\rm d}}
\newcommand{\upsi}{{$\Upsilon$}}

\begin{document}

\begingroup
\thispagestyle{empty}
\baselineskip=14pt
\parskip 0pt plus 5pt

\begin{center}
{\large EUROPEAN LABORATORY FOR PARTICLE PHYSICS}
\end{center}

\bigskip
\begin{flushright}
CERN--PH--EP\,/\,2006--005\\
March 9, 2006
\end{flushright}

\bigskip\bigskip
\begin{center}
\Large\bf{Bottomonium and Drell-Yan production\\ 
in \mbox{p-A} collisions at 450 GeV} 
\end{center}

\bigskip\bigskip
\begin{center}
\emph{\large NA50 Collaboration}
\end{center}
\begin{center}
B.~Alessandro$^{10}$,
C.~Alexa$^{3}$,
R.~Arnaldi$^{10}$,
M.~Atayan$^{12}$,
S.~Beol\`e$^{10}$,
V.~Boldea$^{3}$,
P.~Bordalo$^{6,a}$,
G.~Borges$^{6}$,
J.~Castor$^{2}$,
B.~Chaurand$^{9}$,
B.~Cheynis$^{11}$,
E.~Chiavassa$^{10}$,
C.~Cical\`o$^{4}$,
M.P.~Comets$^{8}$,
S.~Constantinescu$^{3}$,
P.~Cortese$^{1}$,
A.~De~Falco$^{4}$,
N.~De~Marco$^{10}$,
G.~Dellacasa$^{1}$,
A.~Devaux$^{2}$,
S.~Dita$^{3}$,
J.~Fargeix$^{2}$,
P.~Force$^{2}$,
M.~Gallio$^{10}$,
C.~Gerschel$^{8}$,
P.~Giubellino$^{10}$,
M.B.~Golubeva$^{7}$,
A.~Grigoryan$^{12}$,
J.Y.~Grossiord$^{11}$,
F.F.~Guber$^{7}$,
A.~Guichard$^{11}$,
H.~Gulkanyan$^{12}$,
M.~Idzik$^{10,b}$,
D.~Jouan$^{8}$,
T.L.~Karavicheva$^{7}$,
L.~Kluberg$^{9}$,
A.B.~Kurepin$^{7}$,
Y.~Le~Bornec$^{8}$,
C.~Louren\c co$^{5}$,
M.~Mac~Cormick$^{8}$,
A.~Marzari-Chiesa$^{10}$,
M.~Masera$^{10}$,
A.~Masoni$^{4}$,
M.~Monteno$^{10}$,
A.~Musso$^{10}$,
P.~Petiau$^{9}$,
A.~Piccotti$^{10}$,
J.R.~Pizzi$^{11}$,
F.~Prino$^{10}$,
G.~Puddu$^{4}$,
C.~Quintans$^{6}$,
L.~Ramello$^{1}$,
S.~Ramos$^{6,a}$,
L.~Riccati$^{10}$,
H.~Santos$^{6}$,
P.~Saturnini$^{2}$,
E.~Scomparin$^{10}$,
S.~Serci$^{4}$,
R.~Shahoyan$^{6,c}$,
M.~Sitta$^{1}$,
P.~Sonderegger$^{5,a}$,
X.~Tarrago$^{8}$,
N.S.~Topilskaya$^{7}$,
G.L.~Usai$^{4}$,
E.~Vercellin$^{10}$,
N.~Willis$^{8}$.
\end{center}
 
\vfill

\begin{center}
\emph{to be published in Physics Letters B}
\end{center}

\newpage
\begin{center}
\textbf{Abstract}
\end{center}

\bigskip
\begingroup
\leftskip=0.4cm
\rightskip=0.4cm
\parindent=0.pt
\small

The NA50 Collaboration has measured heavy-quarkonium production in \mbox{p-A} 
collisions at 450 GeV incident energy ($\sqrt{s}$ = 29.1 GeV). We report here
results on the production of the $\Upsilon$ states and of high-mass 
Drell-Yan muon pairs (m$_{\mu\mu} >$ 6 GeV/c$^2$). 
The cross-section at midrapidity and the A-dependence of the 
measured yields are determined and compared with the results of other 
fixed-target experiments and with the available theoretical estimates. 
Finally, we also address some issues concerning the transverse momentum
distributions of the measured dimuons.
\endgroup
\bigskip

\setlength{\parindent}{0mm}
\setlength{\parskip}{0mm}
\small
\HRule\\
\begin{flushleft}
$^{~1}$ Universit\`a del Piemonte Orientale, Alessandria and INFN-Torino, 
Italy  \\
$^{~2}$ LPC, Univ. Blaise Pascal and CNRS-IN2P3, Aubi\`ere, France\\
$^{~3}$ IFA, Bucharest, Romania\\
$^{~4}$ Universit\`a di Cagliari/INFN, Cagliari, Italy\\
$^{~5}$ CERN, Geneva, Switzerland\\
$^{~6}$ LIP, Lisbon, Portugal\\
$^{~7}$ INR, Moscow, Russia\\
$^{~8}$ IPN, Univ. de Paris-Sud and CNRS-IN2P3, Orsay, France\\
$^{~9}$ LLR, Ecole Polytechnique and CNRS-IN2P3, Palaiseau, France\\
$^{10}$ Universit\`a di Torino/INFN, Torino, Italy\\
$^{11}$ IPN, Univ. Claude Bernard Lyon-I and CNRS-IN2P3, Villeurbanne, France\\
$^{12}$ YerPhI, Yerevan, Armenia\\

a) also at IST, Universidade T\'ecnica de Lisboa, Lisbon, Portugal\\
b) also at Faculty of Physics and Nuclear Techniques, AGH University of
Science and Technology, Cracow, Poland\\
c) on leave of absence of YerPhI, Yerevan, Armenia\\

\end{flushleft}
\endgroup

\newpage
\pagenumbering{arabic}
\setcounter{page}{1}

 
\section{Introduction}

The study of heavy quarkonium states is very important for our understanding of 
the physics of strong interactions and represents today one of the most 
challenging fields of application of QCD~\cite{QWG04}. 
By studying quarkonium production in \mbox{p-A} collisions at fixed-target 
energies, various aspects of the theory of strong interactions can be addressed. 
In particular, such data can be used to constrain theoretical approaches used 
for the calculation of the production cross-section in nucleon-nucleon
interactions, 
such as the color-evaporation model (CEM)~\cite{Bar80} or the nonrelativistic 
QCD formulation (NRQCD)~\cite{Bod95}. Furthermore, by 
comparing the production data on various nuclear targets, the details of the 
color neutralization process of the $q\overline q$ pair can be 
investigated~\cite{Kop01,Arl00,Vog00}. 
Finally, \mbox{p-A} data are an essential reference for the study of the 
anomalous suppression of quarkonia production predicted to occur 
in ultrarelativistic heavy-ion collisions~\cite{Mat86,Abr00,Ale05}.

To investigate these topics, the NA50 Collaboration performed a systematic 
study of quarkonium production on several nuclear targets with proton beams at 
450 GeV/c incident momentum. Results on the charmonium states \jpsi\ and \psip, 
detected through their decay into two muons, have already been
published~\cite{Ale03,Ale04}. 
Thanks to the large integrated luminosity ($\sim$10 pb$^{-1}$ for each target),
the dimuon mass spectrum also shows a signal 
corresponding to the sum of the \upsi(1S), \upsi(2S) and \upsi(3S) states 
(simply denoted as \upsi\ in the following). 

Various results on \upsi\ production in proton-induced collisions exist at fixed target 
and at ISR energies~\cite{Her77,Yoh78,Uen79,Bad79,Ang79,Kou80,Ant80,Chi85,Yos89,Mor91,
Ald91,Ale96,Bro01,Ned04}, but they are usually less precise than the 
corresponding charmonium results. Up to now, no data are available 
at 450 GeV ($\sqrt{s}$ = 29.1 GeV), and the A-dependence of the production
cross-section has only been studied by the E772 experiment~\cite{Ald91}, 
at a higher incident energy. 

In this Letter we show results on the production of the \upsi\ in proton-induced
collisions on five nuclear targets (Be, Al, Cu, Ag, W). After applying the appropriate cuts, 
the total number of \upsi\ meson events is of the order of 300. We compute, for the
various targets, the cross-sections around midrapidity for the \upsi,
as well as for Drell-Yan muon pairs with invariant mass larger than 6 GeV/c$^2$.
We parametrize these cross-sections with the usual relation 
$\sigma_A\, =\, \sigma_N A^{\alpha}$, 
and determine the values $\alpha_\Upsilon$ and $\alpha_{DY}$. 
We also determine the A-dependence of the ratio 
$B_{\mu\mu}\sigma_{\Upsilon}/\sigma_{DY}$, a quantity less affected by 
systematical effects.
Then, by combining the results for the different target nuclei, we obtain 
$B_{\mu\mu}d\sigma/dy_{cm}|_{y_{cm}=0}$, a quantity that can be compared with the 
results of other experiments and with theoretical calculations. 
Finally, the dependences of $\langle p_{\rm T} \rangle$ and $\langle p_{\rm T}^2 \rangle$ on
$A$ and $\sqrt{s}$ are discussed.

\section{Experimental setup and data analysis}

The setup of the NA50 experiment has been described in detail in various 
publications~\cite{Abr97}. It is based on a muon spectrometer composed of a 5 
meter long hadron absorber, an air-core toroidal magnet and various sets of 
MWPCs and scintillator hodoscopes. 
The luminosity measurement is achieved by means of
three argon ionisation chambers placed along the beam path. They have been 
calibrated at low beam intensity and their linearity has been checked up to 
10$^{10}$ p/s~\cite{Abr98}. 
The intensity of the 450 GeV proton beam delivered by the SPS was about 
3$\cdot$10$^9$ p/burst, with a 2.3 s spill. The thickness of the targets 
ranged from 30 to 50\% of an interaction length.
In these conditions, the dimuon trigger rate was $\sim$10$^3$/burst, 
with a dead time smaller than 5\%. 
The efficiency $\epsilon_{\rm trig}$ of the dimuon trigger system has been
measured during data taking, using dedicated hardware, and ranges from 87 to
89\%~(see Table~\ref{tab:1}).
About 50\% of the triggered events
contain a dimuon, which is reconstructed with the 
improved version of the reconstruction algorithm described in
Refs.~\cite{Ale04,Sha01}. The dimuon reconstruction efficiency, 
$\epsilon_{\mu\mu}$, is dominated by the efficiency of the MWPCs and is 
determined by the reconstruction program.
Its precise value, used as input to obtain the cross-sections, depends on the 
data taking conditions and ranges from 84 to 87\%~(see Table~\ref{tab:1}).
Various quality cuts are applied, to remove parasitic off-target events. 
Furthermore, in order to 
discard dimuons produced at the edge of the
acceptance of the spectrometer, the kinematical domain of the analysis is restricted 
to $-0.5< y_{\rm cm}<0.5$ in rapidity and 
to $-0.5< \cos\theta_{\rm CS}<0.5$ in the
Collins-Soper polar angle of the muons~\cite{Col77}. 
We are finally left with about 
1000 high mass events ($m_{\mu\mu}\,>\, 6$ GeV/c$^2$) for each target.  

\begin{figure}[ht]
\centering
\resizebox{1.0\textwidth}{!}{%
\includegraphics{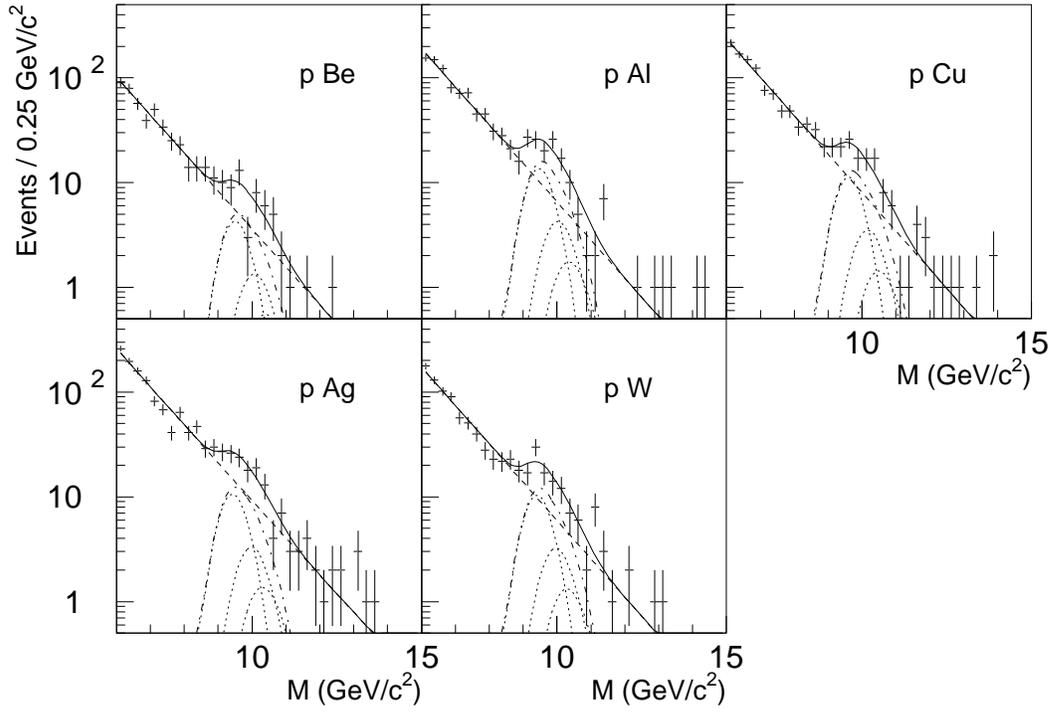}}
\caption{The five \mbox{p-A} opposite-sign muon pair mass spectra, in the mass range
$m_{\mu\mu}>\,$6 GeV/c$^2$. The solid line is the result of the fit with
the function described in Eq.~\ref{eq:ffit}. The dashed line represents the
Drell-Yan process, the dotted lines the 
contributions of the various bottomonium states, the dot-dashed line the sum of the
three \upsi\ states.}
\label{fig:1}
\end{figure}

Figure~\ref{fig:1} shows the five invariant mass spectra, in the region 
$m_{\mu\mu}>6$~GeV/c$^2$. The \upsi\ signal can be clearly seen on top of a 
continuum due to the Drell-Yan process. In this mass region the combinatorial
background due to $\pi$ and $K$ decays, usually calculated through the measured 
like-sign dimuon sample, turns out to be negligible.
Muon pairs originating from Drell-Yan and from the \upsi\ resonances
are estimated by means of a fit to the invariant mass spectrum.
The mass shapes for the various processes have been calculated through a 
Monte-Carlo simulation. The Drell-Yan events have been generated using a LO 
QCD calculation, with the GRV94LO set of parton distribution
functions~\cite{Glu95};
the $p_{\rm T}$ distributions have been tuned directly on the data (see Section~\ref{sec:3} for
details) and, in the
considered mass range, they do not significantly depend on the dimuon mass.  
The bottomonium 
states have been generated at their nominal PDG mass~\cite{PDG04}. 
Due to the invariant mass resolution of the spectrometer, of the order 
of 0.4 GeV/c$^{2}$ for dimuons with $m_{\mu\mu}\sim$ 10 GeV/c$^2$, the various \upsi\ 
states cannot be resolved. 
We have, therefore, relied on previous measurements to fix the relative 
contributions of the three states visible in the dimuon channel. 
We have used the values $\Upsilon(2\rm S)/\Upsilon(1\rm S)=0.32$ and 
$\Upsilon(3\rm S)/\Upsilon(1\rm S)=0.13$, 
as measured in p-A collisions at 400 GeV~\cite{Uen79}, i.e.\ close to our 
energy. The results of Ref.~\cite{Uen79} are anyway compatible with a high 
statistics measurement performed at higher energy~\cite{Mor91}. 
For the bottomonium $p_{\rm T}$ distribution, we have adopted the 
functional form
\begin{equation}
{\rm d}\sigma/{\rm d}p_{\rm T}\propto p_{\rm T}/\left[1+\left(p_{\rm T}/p_{0}\right)^{2}\right]^{6}, 
\label{eq:pt}
\end{equation}
used in Ref.~\cite{Mor91}. The $p_0$ parameter has been fitted on the measured 
$p_{\rm T}$ distributions (see Section~\ref{sec:3}). 

For the rapidity, due to the narrow coverage of the set-up
and to the relatively small statistics, our data do not allow a precise determination 
of the \upsi\ $y$-distribution. As a first guess, we have used in the acceptance
calculation a gaussian distribution, centered at $y_{\rm cm}$=0, with
$\sigma_{\rm y}=0.37\pm 0.01$. This value comes from a fit to the $x_{\rm F}$
distribution for \upsi\ production at 400 GeV ($\sigma_{\rm y}=0.35\pm 0.01$)~\cite{Uen79,Chi85}. The obtained value
has then been logarithmically scaled to our incident energy.
In Section~\ref{sec:3} we will investigate in more detail the influence
of the rapidity shape on our \upsi\ results. 
Finally, the $\cos\theta_{\rm CS}$ distributions for the \upsi\ states have been
generated according to the recent results of the E866 
Collaboration~\cite{Bro01}, which 
found a negligible polarization for the \upsi(1S) around midrapidity and a 
large transverse polarization for the \upsi(2S) and \upsi(3S) states.
The generated Drell-Yan and \upsi\ events have then been tracked through the 
apparatus and the accepted events have been reconstructed using the
procedure applied to the experimental data.
With the choice of the generation parameters described above, the acceptances 
for the various dimuon sources turn out to be 
$A_{DY}\, =\, 21$\% ($m_{\mu\mu}>\,6$ GeV/c$^2$), 
$A_{\Upsilon(1\rm S)}\, = \, 25$\%, $A_{\Upsilon(2\rm S)}\, = \, 25$\%, 
$A_{\Upsilon(3\rm S)}\, = \, 26$\%.
These acceptances refer to the phase space domain $\mathcal D$ defined by the 
cuts $-0.5\,<y_{\rm cm}<\,0.5$ and $-0.5<\,\cos\theta_{\rm CS}<\,0.5$.
The error on the acceptances due to the uncertainty on the rapidity distribution is less than 1 \%.

Having calculated the shapes of the expected contributions, we fit each of the 
five invariant mass spectra with
the function

\begin{equation}
\frac{\dd N^{+-}}{\dd M}~=~ N^{DY}\frac{\dd N^{DY}}{\dd M}
+N^{\Upsilon}
\left(\frac{\dd N^{\Upsilon(1\rm S)}}{\dd M} + 
0.32\cdot\frac{A_{\Upsilon(2\rm S)}}{A_{\Upsilon(1\rm S)}} \frac{\dd N^{\Upsilon(2\rm S)}}{\dd M} +
0.13\cdot\frac{A_{\Upsilon(3\rm S)}}{A_{\Upsilon(1\rm S)}} \frac{\dd N^{\Upsilon(3\rm S)}}{\dd M}
\right) 
\label{eq:ffit}
\end{equation}

\noindent where $N^{DY}$ and $N^{\Upsilon}$ are free parameters in the fit.
To compensate for possible approximations in the description of the setup and 
in the mapping of the magnetic field we leave the position of the \upsi(1S) peak 
as a further free parameter in the fit. The mass differences 
$m_{\Upsilon(2S)}-m_{\Upsilon(1S)}$ and $m_{\Upsilon(3S)}-m_{\Upsilon(1S)}$ are 
anyway kept fixed to their PDG values~\cite{PDG04}.
We find that the shift from the nominal mass of the bottomonium states does not 
exceed 1\% and has negligible consequences ($<$ 1\%) on the estimated \upsi\ 
yield.

The choice of the the fit region is dictated by the 
request of having a negligible contribution of the charmonia states. 
To satisfy this request, one could start the fit at mass 
values around 4.5 GeV/c$^2$. However, due to the steep slope of the
Drell-Yan mass distributions, the use of a large fit window 
results in a loss of weight of the low-statistics \upsi\ region.
In this way, small inaccuracies in the Monte-Carlo description of the 
Drell-Yan might influence the estimate of the \upsi\ yield.
Pushing down to 4.5 GeV/c$^2$ the starting point of 
the fit, the change in the calculated number of Drell-Yan and \upsi\ events is 
only 10\% on average, but the $\chi^2$ of the fits are worsened by about 50\%.
Therefore, we have started our fits at $m_{\mu\mu}=\,6$ GeV/c$^2$. 
In this way, as can be seen in Fig.~\ref{fig:1}, the measured invariant mass 
spectra are well reproduced, with values of $\chi^2/ndf$ ranging from 
0.9 to 1.5.
The number of \upsi\ mesons and the number of high mass Drell-Yan pairs
are reported in Table~\ref{tab:1}.

Finally, the Drell-Yan and \upsi\ cross-sections, in the kinematical domain
$\mathcal D$ defined above, have been calculated through the formula

\begin{equation}
\sigma^i_{\mathcal D}=
\frac{N_i}{A_i}\cdot\frac{1}{{\mathcal L}\cdot 
\epsilon_{\mu\mu}\cdot\epsilon_{\rm trig}} \quad ,
\label{eq:cross}
\end{equation}
\noindent
where $N_i$ is the number of detected 
events for each process $i$, $\mathcal L$ is the luminosity, corrected for dead 
time, $A_i$ is the acceptance, $\epsilon_{\mu\mu}$ is the dimuon reconstruction 
efficiency and $\epsilon_{\rm trig}$ is the efficiency of the
dimuon trigger. For the \upsi, we calculate the cross-section integrated over 
the $\Upsilon$(1S), $\Upsilon$(2S) and $\Upsilon$(3S) states. 
The values of the various quantities relevant for the determination of the  
cross-section are summarized in Table~\ref{tab:1}.

\section{Results}
\label{sec:3}

\begin{figure}[ht]
\centering
\resizebox{1.0\textwidth}{!}{%
\includegraphics{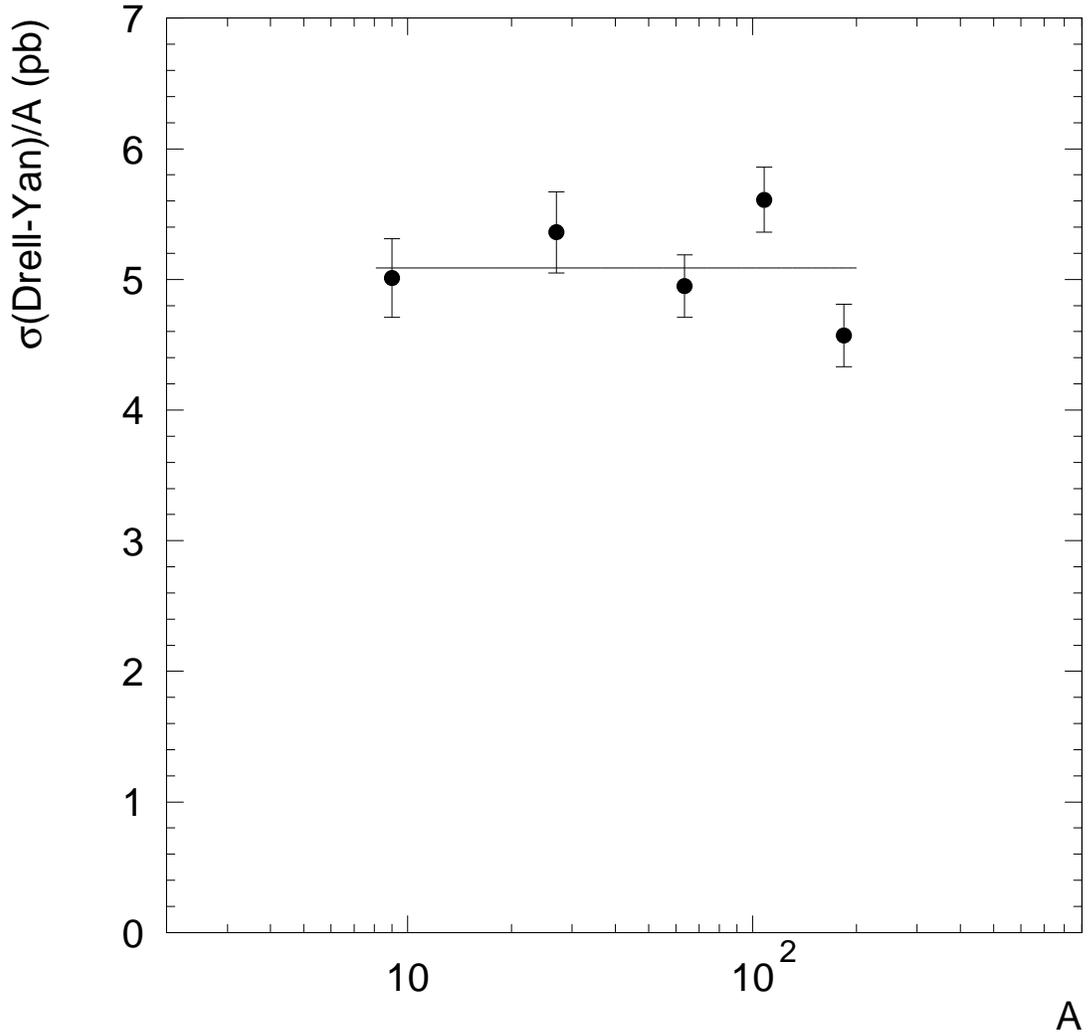}}
\caption{The Drell-Yan cross-section, relative to the mass region 
$m_{\mu\mu}>\,$6 GeV/c$^2$, divided by the mass number $A$. The line represents
the result of a fit to the points, according to the function 
$\sigma_{DY}^{pA}\,=\,\sigma_{DY}^{pp}\cdot A$.}
\label{fig:2}
\end{figure}

In Fig.~\ref{fig:2} we show, as a function of the mass number $A$, the 
Drell-Yan cross-sections divided by $A$. The points refer to the kinematical domain 
$\mathcal D$, and to the mass range $m_{\mu\mu}>$ 6 GeV/c$^2$. 
The error bars shown in Fig.~\ref{fig:2} 
represent the quadratic combination of statistical and systematical errors. The
systematical errors on the cross-section measurements are due to the errors on 
the determination of the incident proton flux and on the evaluation of 
$\epsilon_{\rm trig}$ and $\epsilon_{\mu\mu}$~\cite{Ale04}. 
They range from 3.4 to 3.7\% and 
are comparable to the statistical errors.
The quantity $\sigma_{DY}/A$ corresponds to the cross-section per 
nucleon-nucleon collision. In absence of final state interactions, and if
nuclear effects on the PDFs are negligible, a flat behaviour is expected. 
By fitting the points with the function 

\begin{equation}
\sigma_{DY}^{pA} = \sigma_{DY}^{pp}\cdot A^{\alpha_{DY}}
\label{eq:funcdy}
\end{equation}

\noindent we get $\alpha_{DY}$=0.98$\pm$0.02 ($\chi^2/ndf$ = 3.1), a value 
compatible with 1.
Alternatively, imposing $\alpha$=1 results in the line shown in 
Fig.~\ref{fig:2}, with $\chi^2/ndf$ = 2.5. This result is in fair agreement with 
previous observations by NA50 in a lower mass range~\cite{Ale03,Ale04}, showing that 
the Drell-Yan cross-section scales with the number of nucleon-nucleon collisions. 
 
\begin{figure}[ht]
\centering
\resizebox{1.0\textwidth}{!}{%
\includegraphics{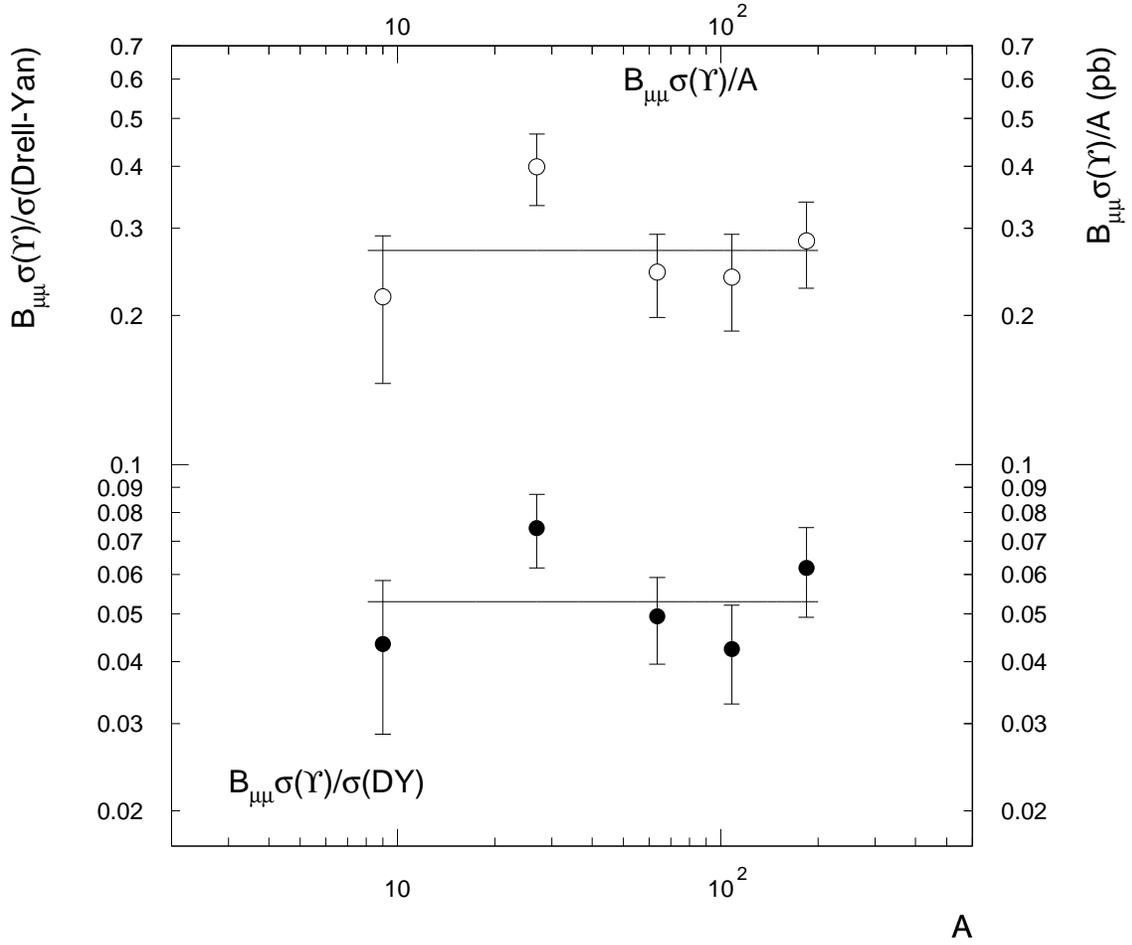}}
\caption{The ratio $B_{\mu\mu}\sigma_{\Upsilon}/\sigma_{DY}$ (closed circles)
and the cross-section per nucleon
 $B_{\mu\mu}\sigma_{\Upsilon}/A$ (open circles), as a function of $A$. 
}
\label{fig:3}
\end{figure}

Figure~\ref{fig:3} shows the A-dependence of bottomonium production,
through the ratio 
$B_{\mu\mu}\sigma_{\Upsilon}/\sigma_{DY}$, where
$B_{\mu\mu}\sigma_{\Upsilon}$ refers to the sum of the 1S, 2S and 3S states, 
weighted by their branching ratio into two muons. The quantity 
$B_{\mu\mu}\sigma_{\Upsilon}/\sigma_{DY}$ is proportional to the 
bottomonium cross-section per nucleon-nucleon collision. Being the
ratio of two measured cross-sections it is less affected by  
systematical errors.
In the same figure we show the quantity 
$B_{\mu\mu}\sigma_{\Upsilon}/A$ as a function of $A$. Again, the plotted errors 
represent a quadratic combination of statistical and systematical errors. 
The values of the various cross-sections and of their ratios is shown in
Table~\ref{tab:2}.

The fits of the 
two sets of points to the usual $A^{\alpha}$ parameterization give
$\alpha_{\Upsilon/DY}\,=\,0.98\pm 0.10$ ($\chi^2/ndf$=1.7) and 
$\alpha_{\Upsilon}\,=\,0.98\pm 0.10$ ($\chi^2/ndf$=1.7). Within the rather large
statistical errors, our result indicates that the \upsi\ 
is not strongly absorbed 
in the 
nuclear medium. In particular, by imposing $\alpha$ = 1 (solid lines in Fig.~\ref{fig:3}),
 we can describe the 
data with $\chi^2/ndf$ = 1.3 for \upsi/DY and for $\sigma_{\Upsilon}$.
We find that our result is compatible with the only other direct determination of 
the $A$-dependence of \upsi\ production, carried out by E772~\cite{Ald91} at
800 GeV incident energy. 

\begin{figure}[ht]
\centering
\resizebox{1.0\textwidth}{!}{%
\includegraphics{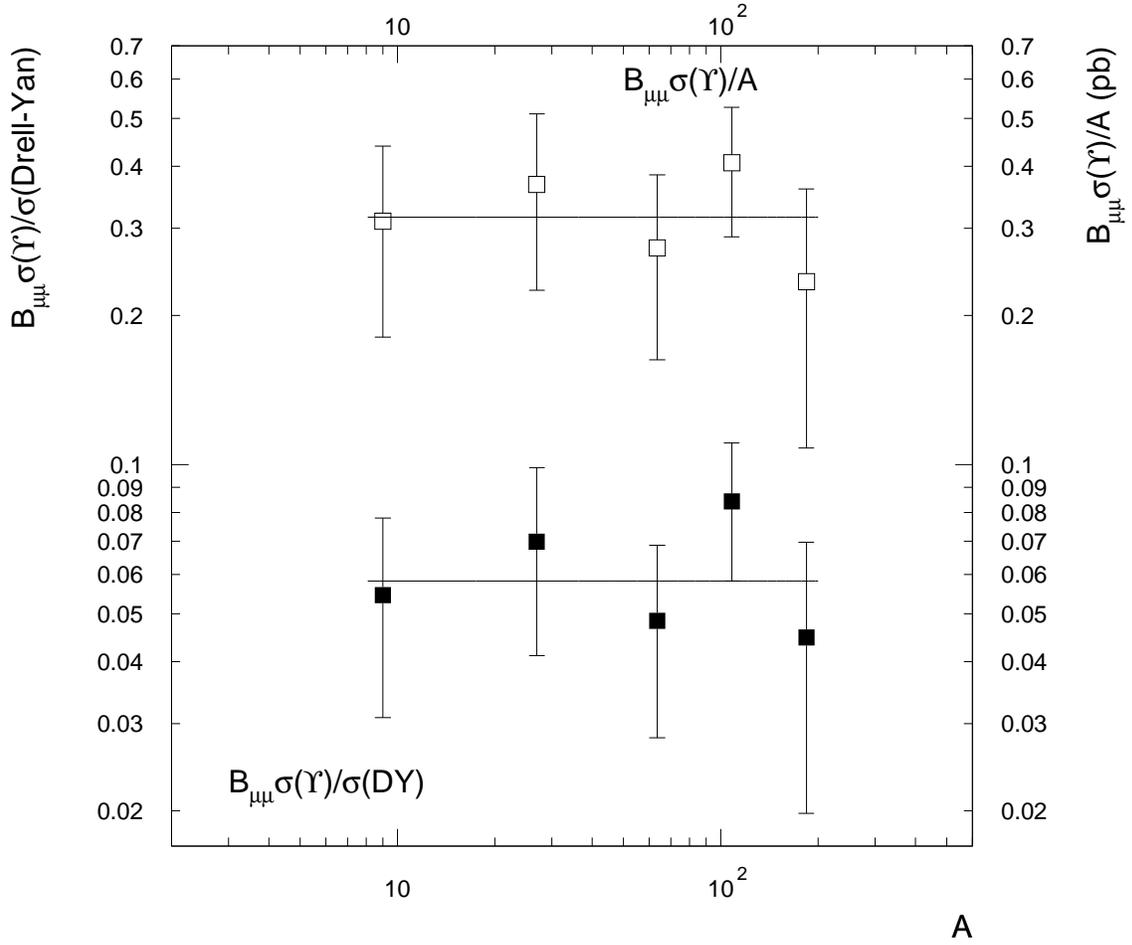}}
\caption{The ratio $B_{\mu\mu}\sigma_{\Upsilon}/\sigma_{DY}$ (closed squares)
and the cross-section per nucleon 
$B_{\mu\mu}\sigma_{\Upsilon}/A$ (open squares), as a function of $A$, for the 
low-intensity data sample. 
}
\label{fig:3b}
\end{figure}

Assuming $\alpha_{\Upsilon}\,=\,1$, as suggested by our data, we averaged the 
results on the various targets and obtained for the \upsi\ cross-section per 
nucleon-nucleon collision the value
$B_{\mu\mu}\sigma_{\Upsilon}\,=\,0.27\pm0.03$ pb/nucleon. As a further
check of possible systematic effects, we also obtained the same
quantity through the analysis of a different
data sample, taken by NA50~\cite{Ale03} at 
a beam intensity about one order of magnitude lower and containing
72$\pm$12 \upsi\ events. We get 
$B_{\mu\mu}\sigma_{\Upsilon}\,=\,0.32\pm0.06$ pb/nucleon.
The two values of $\sigma_{\Upsilon}$ agree within errors, 
indicating that the high-luminosity data analyzed in this paper
have been properly corrected for the various efficiency factors,
whose evaluation is more delicate at high beam intensities.
For completeness, we show in Fig.~\ref{fig:3b}, as a function of $A$, 
$B_{\mu\mu}\sigma_{\Upsilon}/\sigma_{DY}$ and 
$B_{\mu\mu}\sigma_{\Upsilon}$ for the low beam intensity sample.  
Performing a simultaneous fit of the data sets of Fig.~\ref{fig:3} 
and~\ref{fig:3b}, we get $\alpha_{\Upsilon/DY}\,=\,0.98\pm 0.09$ 
($\chi^2/ndf$=0.9) and $\alpha_{\Upsilon}\,=\,0.98\pm 0.08$ ($\chi^2/ndf$=0.8).
 
The \upsi\ cross-section per nucleon-nucleon collision can be compared with 
previous measurements performed by various fixed-target 
experiments. However, past experiments usually quote the quantity 
$B_{\mu\mu}{\rm d}\sigma/{\rm d}y_{cm}|_{y_{cm}=0}$, 
and assumptions different from ours have sometimes been adopted in the
cross-section calculations.
In particular, past experiments evaluated their acceptances 
assuming an unpolarized production of the \upsi\ states. 
Furthermore, in
some cases they applied a correction to take into account the Fermi-motion of the 
target nucleons. In order to properly compare our results with the 
ones available in the literature, we have first re-calculated our acceptances 
assuming no polarization 
for all the three bottomonium states. With respect 
to the results shown in Fig.~\ref{fig:3} and~\ref{fig:3b} we get a 1.5\% decrease in the value 
of the cross-section. The influence of Fermi motion has been taken into account
by applying to our results (and to the ones available in the literature not yet corrected) 
the correction proposed in Ref.~\cite{Yoh78}.
%
%
At our center of mass energy, $\sqrt{s}$ = 29.1 GeV, the factor to apply to the 
results is 0.897. 

Concerning the $A$-dependence, all the other experiments also assumed $\alpha$=1.

A further element in the calculation of the midrapidity cross-section is the
rapidity shape assumed for the \upsi\ production. In fact, due to the 
limited y-coverage of the NA50 setup, our result may be sensitive to the 
chosen rapidity shape. 
We find that the rapidity distribution of the \upsi\ mesons in our sample
is compatible with a gaussian having $\sigma_y^{\Upsilon}>$ 0.30 
(95\% c.l.).  Indeed, it is even compatible with a flat y-distribution.
The value for 
$B_{\mu\mu}{\rm d}\sigma/{\rm d}y_{cm}|_{y_{cm}=0}$ varies by $^{+7\%}_{-15\%}$ in these two 
extreme cases, and we consider this uncertainty as a contribution to the systematical
error in the determination of the mid-rapidity cross-section. 

\begin{figure}[ht]
\centering
\resizebox{1.0\textwidth}{!}{%
\includegraphics{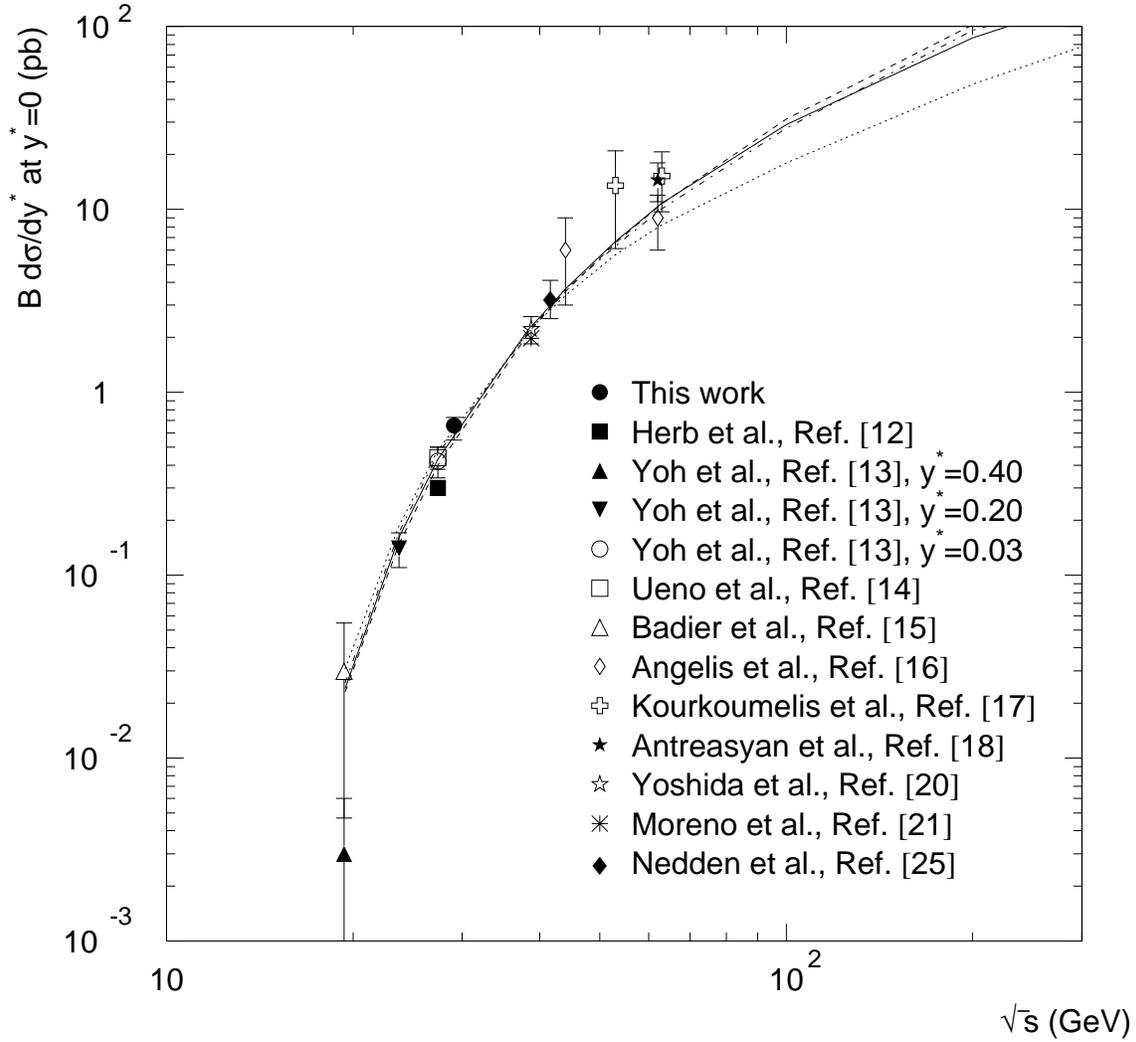}}
\caption{\upsi\ cross-section at midrapidity as a function of $\sqrt{s}$. The
lines represent the results of NLO CEM calculations~\cite{Vog99}. 
The solid, dashed and dot-dashed lines are obtained with the MRST HO
PDF set with, respectively,  $m_{b}\,=\,\mu\,=\,4.75$~GeV/c$^2$, 
$m_{b}\,=\,\mu/2\,=\,4.5$~GeV/c$^2$, and $m_{b}\,=\,2\mu\,=\,5$~GeV/c$^2$.
The dotted line is obtained with the GRV HO PDF set with $m_{b}\,=\,\mu\,=\,4.75$~GeV/c$^2$. 
It is assumed that
$\mu=\mu_{R}=\mu_{F}$, where $\mu_{R}$ is the renormalization scale and 
$\mu_{F}$ is the factorization scale.}
\label{fig:4}
\end{figure}

The obtained value is $B_{\mu\mu}{\rm d}\sigma/
{\rm d}y|_{y_{cm}=0}\,=\,0.65\pm0.05^{+0.04}_{-0.10}$~pb/nucleon, where 
the quoted asymmetric error is due to the uncertainty on the \upsi\ rapidity distribution 
as derived from our data.
In Fig.~\ref{fig:4} we
present the $\sqrt{s}$-dependence of the \upsi\ cross-section per 
nucleon-nucleon collision.
The NA50 point follows the general trend defined by previously available results.
We also plot in Fig.~\ref{fig:4} the results of a NLO
calculation based on the Color Evaporation Model (CEM)~\cite{Vog99}. 
The curves are obtained by fitting the data with 
various combinations of the CEM parameters, namely the mass of the b-quark, the
renormalization/factorization scale and the choice of the PDF set. 
It can be seen that the $\sqrt{s}$ dependence of the
\upsi\ cross-section is well reproduced. For the other model commonly used in 
the study of heavy quarkonia cross-sections, NRQCD, no up-to-date calculation 
exists in the literature. The only available prediction~\cite{Ben96} gives
a value of the order of 50 pb/nucleon for the total \upsi\ cross-section 
at $\sqrt{s}\,=\,29.1$ GeV.
By integrating our differential cross-section
 over 
$y$ and $\cos \theta_{CS}$, and correcting for the branching ratios, we get 
$\sigma_{\rm tot}^{\Upsilon}$ = 29$\pm 2^{+31}_{-4}$ pb/nucleon. The asymmetric systematic error is
dominated by the uncertainty in the rapidity distribution derived from our data.
If we rely on the data from previous experiments for the determination of $\sigma_{y}^{\Upsilon}$ the systematic
error is reduced to $\pm 2$ pb/nucleon.

\begin{figure}[ht]
\centering
\resizebox{1.0\textwidth}{!}{%
\includegraphics{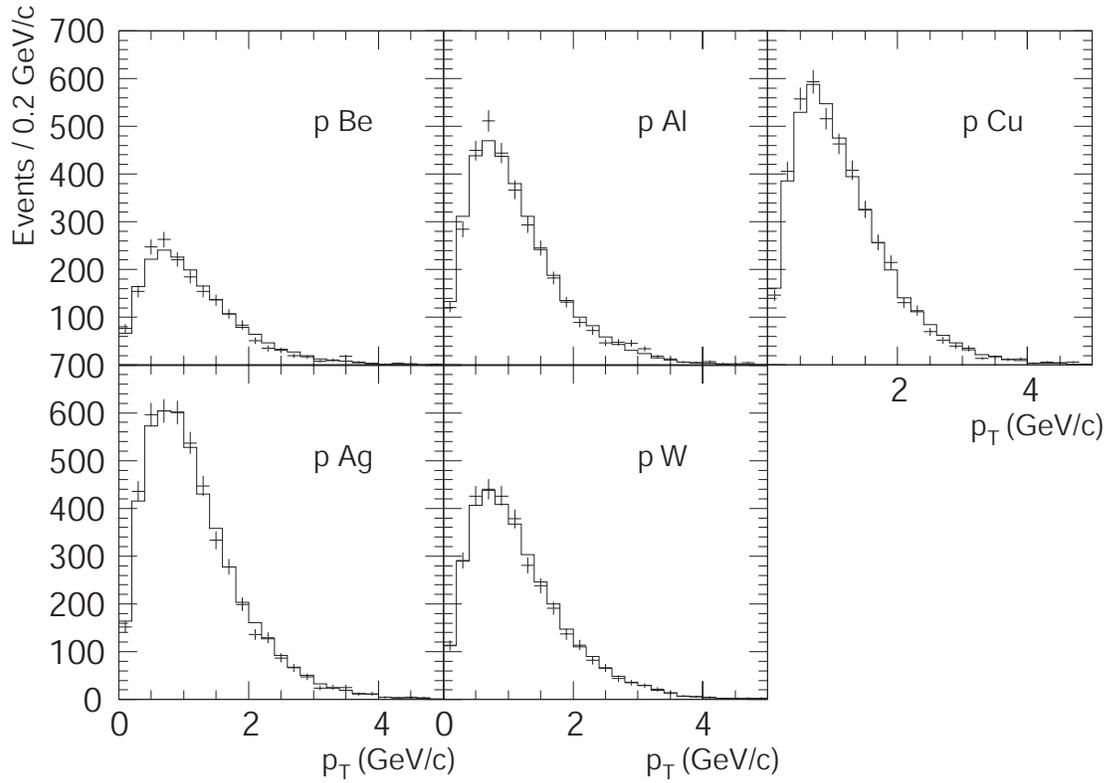}}
\caption{The measured \mbox{p-A} opposite-sign dimuon $p_{\rm{T}}$
spectra in the invariant mass region $4.5<m_{\mu\mu}<8.0$ GeV/c$^{2}$.
The lines represent the best fit to the data obtained with the functional form shown in
Eq.~\ref{eq:pt}.} 
\label{fig:5bis}
\end{figure}

\begin{figure}[ht]
\centering
\resizebox{1.0\textwidth}{!}{%
\includegraphics{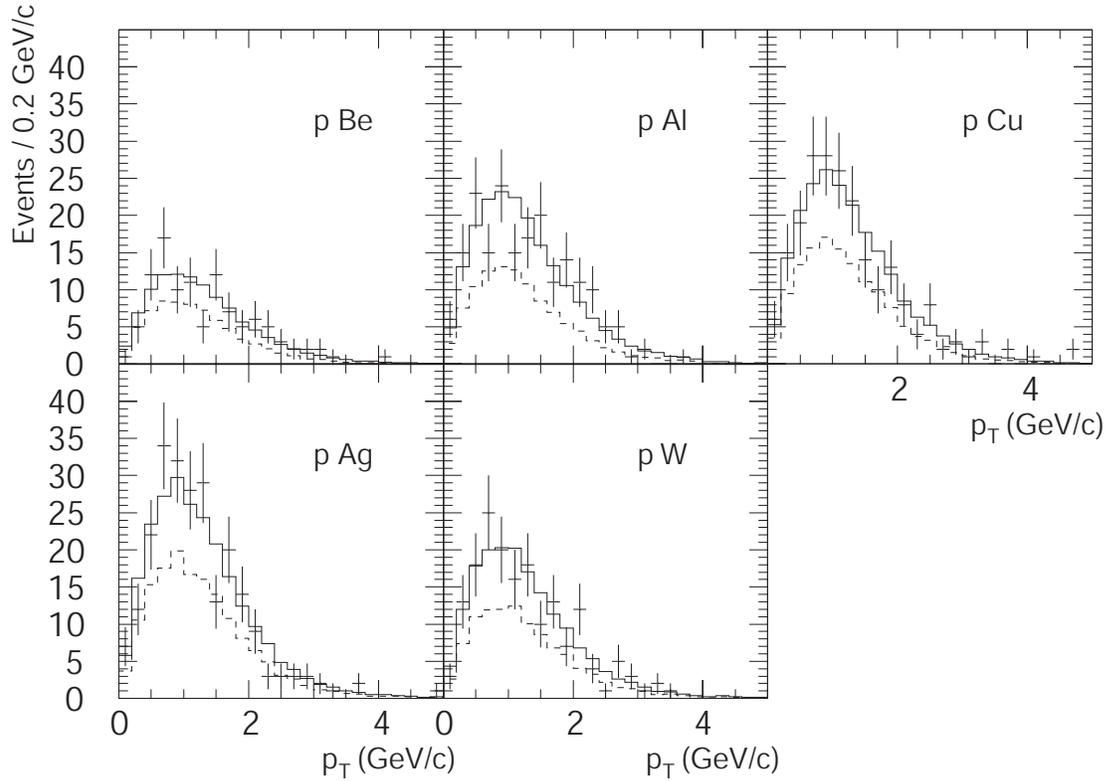}}
\caption{The measured \mbox{p-A} opposite-sign dimuon $p_{\rm{T}}$
spectra in the invariant mass region $8.6<m_{\mu\mu}<11.6$ GeV/c$^{2}$.
The dashed lines represent the contribution of the Drell-Yan process, the solid lines the sum of
the \upsi\ and Drell-Yan yields (see text for details).} 
\label{fig:5tris}
\end{figure}

Finally, we have investigated the $A$-dependence of
the dimuon transverse momentum distributions. In Figs.~\ref{fig:5bis} 
and~\ref{fig:5tris} we show, for the various systems, the $p_{\rm T}$ 
spectra observed in the mass intervals $4.5<\,m_{\mu\mu}\,<8$ GeV/c$^2$ 
and $8.6<\,m_{\mu\mu}\,<11.6$ GeV/c$^2$, respectively.
In Fig.~\ref{fig:5bis}, the Drell-Yan $p_{\rm T}$ distributions, generated 
according to Eq.~\ref{eq:pt} and filtered through the Monte-Carlo simulation of the detector, 
have been fitted to the data with $p_0$ as a free parameter. 
In Fig.~\ref{fig:5tris} the spectra have been fitted with a superposition of 
Drell-Yan and \upsi\ dimuons. For Drell-Yan (dashed lines) the normalizations have been obtained 
from the fit to the mass distributions of Fig.~\ref{fig:1}, while the shapes have been  
extrapolated from the $p_{\rm T}$ spectra measured in the lower mass range. 
The \upsi\ events have been generated following Eq.~\ref{eq:pt}, and $p_0$ has been used as a 
free parameter in the fits. 
As a result, we find that for the two mass intervals under study our fits nicely reproduce the 
observed $p_{\rm T}$ distributions. In Table~\ref{tab:3} we summarize the $p_0$ values obtained for
the various systems, for both Drell-Yan and \upsi. The calculated $\langle p_{\rm T}\rangle$ and
$\langle p^2_{\rm T}\rangle$ are also shown. 
To test the stability of our results on the choice of the assumed Drell-Yan 
$p_{\rm T}$ distributions, we have performed a similar study, fitting our data 
to a PYTHIA calculation,
performed with the GRV94LO set of parton distribution functions. The value of 
the intrinsic $\langle k{_\perp}\rangle$ of partons inside hadrons, used in
PYTHIA, has a strong 
influence on the Drell-Yan $p_{\rm T}$ distribution. It has been tuned 
directly on the data, obtaining values ranging from 0.88 to 0.90 GeV/c.
We find that the shape of the PYTHIA $p_{\rm T}$ distribution still reproduces 
our data, with a slightly worse $\chi^2/ndf$. 
The calculated $\langle p_{\rm T}\rangle$ differ by less than 3\% 
(7\% for $\langle p_{\rm T}^2\rangle$) from the result obtained using 
Eq.~\ref{eq:pt}.
 
\begin{figure}[ht]
\centering
\resizebox{1.0\textwidth}{!}{%
\includegraphics{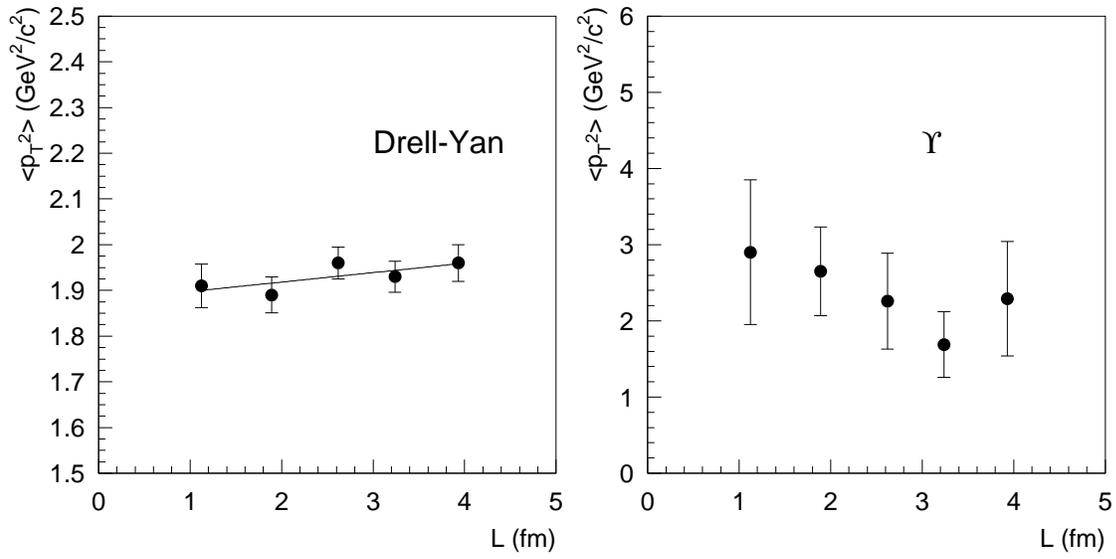}}
\caption{The $L$-dependence of the average transverse momentum squared for
Drell-Yan dimuons in the mass range $4.5\,<m_{\mu\mu}<\,8$ GeV/c$^2$ (left plot) and for \upsi\
(right plot). For Drell-Yan, the result of a linear fit is also shown.}
\label{fig:5}
\end{figure}

\begin{figure}[ht]
\centering
\resizebox{1.0\textwidth}{!}{%
\includegraphics{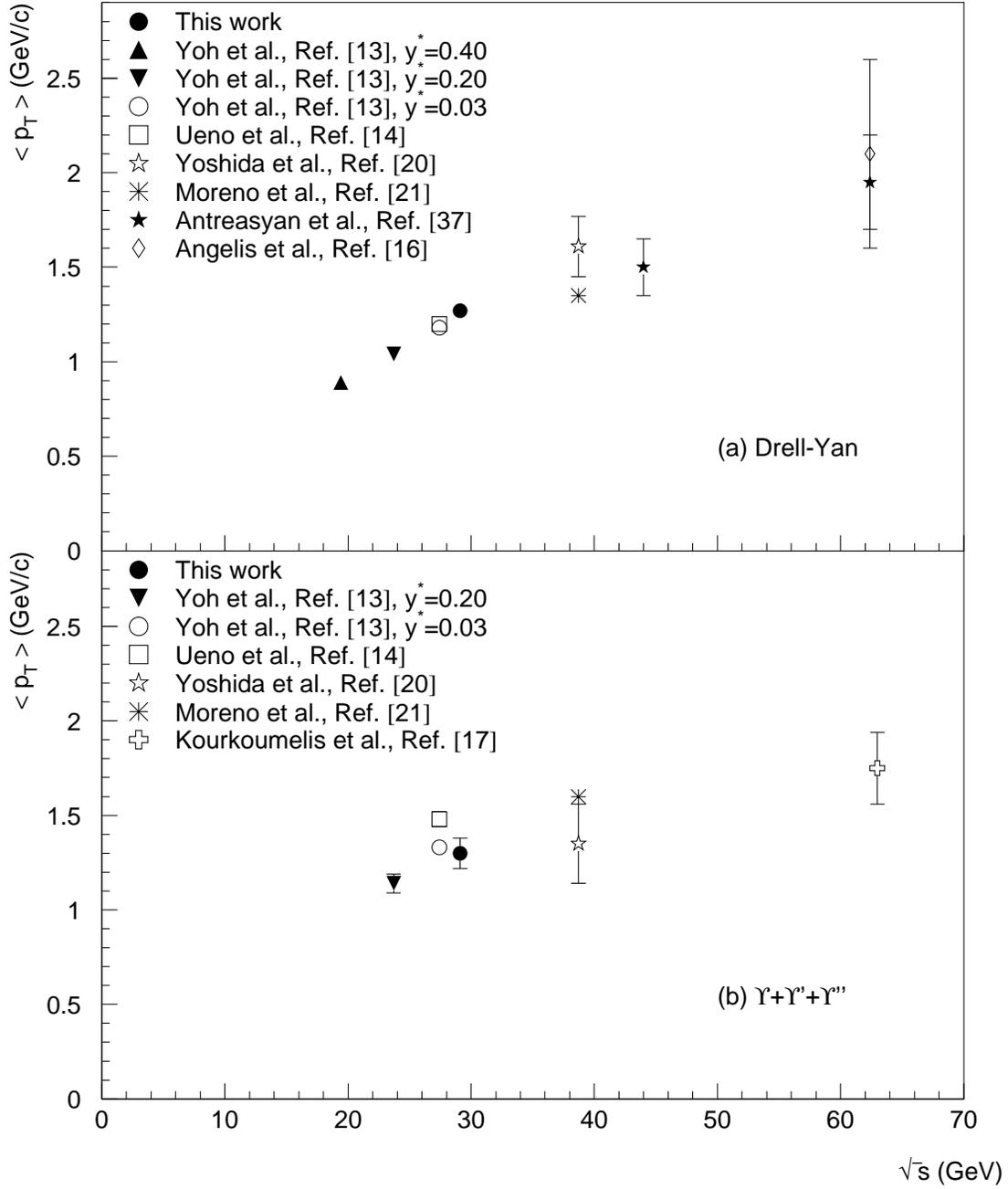}}
\caption{The average transverse momentum for Drell-Yan (a) and \upsi\ (b) production as
a function of $\sqrt{s}$. For Drell-Yan, the points correspond to
$m\sim0.22\cdot\sqrt{s}$ (corresponding to $\sqrt{\tau}\,=\,m/\sqrt{s}\sim0.22$).}
\label{fig:6}
\end{figure}

According 
to~\cite{Huf88,Bla89}, for heavy quarkonia and Drell-Yan production, 
initial state parton scattering is responsible for the increase of 
$\langle p^2_{\rm T}\rangle$ observed in \mbox{p-A} collisions with respect to 
\mbox{pp}. Such an increase is approximately linear as a 
function of $L$, the mean length of nuclear matter crossed by the incoming 
parton, and its size is smaller (by a factor 4/9) if the scattering process is 
initiated by a quark rather than by a gluon.
In Ref.~\cite{Top03} it was shown that the transverse momentum
distribution of the ${\rm J}/\psi$ mesons produced in \mbox{p-A} and \mbox{A-A} collisions
can be fitted with the function $\langle p^2_{\rm T}\rangle = 
\langle p^2_{\rm T}\rangle_{pp} + a_{gN}\cdot L$, with $a_{gN}$ = 
0.077$\pm$0.002 (GeV/c)$^2$fm$^{-1}$.
In Fig.~\ref{fig:5} we show the $L$-dependence of $\langle p^2_{\rm T}\rangle$ for
Drell-Yan dimuons in the mass region $4.5<\,m_{\mu\mu}\,<8$ GeV/c$^2$, where the
initial hard parton is a quark rather than a gluon. A linear increase of  
$\langle p^2_{\rm T}\rangle$ with $L$ is visible. By using the same kind of fit,
we get $a_{qN}$=0.021$\pm$0.019 (GeV/c)$^2$fm$^{-1}$, a value significantly
lower than $a_{gN}$ and, within the rather large errors, in agreement
with the expected 4/9 factor.
For \upsi\ production, where both
$q\overline q$ and $gg$ initial states contribute~\cite{Ben96}, we would expect for the 
$\langle p^2_{\rm T}\rangle$ increase vs.\ $L$ an intermediate slope between the ones
observed for Drell-Yan and $\rm{J}/\psi$. Unfortunately, the available statistics 
(see Fig.~\ref{fig:5}) does not allow to draw quantitative conclusions.
Finally, in Fig.~\ref{fig:6} we compare, as a function of $\sqrt{s}$, 
the measured $\langle p_{\rm T}\rangle$ (averaged over the various target nuclei) 
with the results of other experiments. For Drell-Yan, 
a clear increase of $\langle p_{\rm T}\rangle$ with $\sqrt{s}$ is observed, while 
the effect is less important for the \upsi. 

\section{Conclusions}

The NA50 experiment measured high-mass ($m_{\mu\mu}>\,$6
GeV/c$^2$) Drell-Yan dimuons and, for the first time, \upsi\ production in p-A 
collisions at $\sqrt{s}\,=\,$29.1 GeV. By fitting the
cross-section results with the usual $A^{\alpha}$ parameterization, we get
$\alpha_{DY}\,=\,0.98\pm0.02$ and $\alpha_{\Upsilon}\,=\,0.98\pm0.08$. Within 
the rather large experimental errors, this result indicates the absence of strong absorption 
effects for the bottomonium states in nuclear matter. The measured cross-section at mid-rapidity is in good agreement with the results of other experiments in the fixed 
target energy range and with the available theoretical calculations.
The $A$-dependence of $\langle p^2_{\rm T}\rangle$ indicates, for Drell-Yan production, 
an effect compatible with quark scattering in the initial state.


\begin{table}
\begin{center}{\small }\begin{tabular}{|c|c|c|c|c|c|}
\hline 
&
 N$_{\Upsilon}$&
 N$_{DY}$ (m$_{\mu\mu}>\,6$ GeV/c$^{2}$) &
 $\mathcal{L}$ (pb$^{-1}$) &
 $\epsilon_{\mu\mu}$&
 $\epsilon_{\textrm{trig}}$\tabularnewline
\hline
\mbox{p-Be}&
 26$\pm$9 & 485$\pm$23 & 52.2 & 0.86 & 0.87\tabularnewline
\hline
\mbox{p-Al}&
 82$\pm$13 & 901$\pm$31& 30.2 & 0.87 & 0.88\tabularnewline
\hline
\mbox{p-Cu}&
 67$\pm$13 & 1110$\pm$34 & 17.1 & 0.84 & 0.89\tabularnewline
\hline
\mbox{p-Ag}&
 65$\pm$14 & 1243$\pm$37 & 9.9 & 0.86 & 0.89\tabularnewline
\hline
\mbox{p-W}&
 63$\pm$12 & 820$\pm$30 & 4.7 & 0.84 & 0.87 \tabularnewline
\hline
\end{tabular}\end{center}{\small \par}

\caption{{\label{tab:1}Quantities used in the cross-section calculation.}}
\end{table}
{\small \par}

{\small }%
\begin{table}
\begin{center}{\small }\begin{tabular}{|c|c|c|c|}
\hline 
&
 $\sigma_{DY}$/A (pb)&
 $B_{\mu\mu}\sigma_{\Upsilon}$/A (pb)&
 $B_{\mu\mu}\sigma_{\Upsilon}/\sigma_{DY}$\tabularnewline
\hline
\mbox{p-Be}&
 5.01$\pm$0.24$\pm$0.18 &
 0.218$\pm$0.072$\pm$0.008 &
 0.044$\pm$0.015 \tabularnewline
\hline
\mbox{p-Al}&
 5.36$\pm$0.19$\pm$0.25 &
 0.399$\pm$0.063$\pm$0.019 &
 0.075$\pm$0.012 \tabularnewline
\hline
\mbox{p-Cu}&
 4.95$\pm$0.15$\pm$0.19 &
 0.245$\pm$0.046$\pm$0.009 &
 0.049$\pm$0.010 \tabularnewline
\hline
\mbox{p-Ag}&
 5.61$\pm$0.17$\pm$0.19 &
 0.239$\pm$0.052$\pm$0.008 &
 0.043$\pm$0.009 \tabularnewline
\hline
\mbox{p-W}&
 4.57$\pm$0.17$\pm$0.17 &
 0.283$\pm$0.055$\pm$0.010 &
 0.062$\pm$0.013  \tabularnewline
\hline
\end{tabular}\end{center}{\small \par}

\caption{{\label{tab:2}Drell-Yan (for the mass region $m_{\mu\mu}>\,$6
GeV/c$^{2}$) and bottomonium cross-sections, and their ratio. The first
error is statistical, the second systematical. The quoted systematical
errors cancel out in the ratio $B_{\mu\mu}\sigma_{\Upsilon}/\sigma_{DY}$.}}
\end{table}

\begin{table}[ht]
\centering
\begin{tabular}{|c|c|c|c|c|c|c|}
\hline
 & $p_{0}$ (DY) & $p_{0}$ (\upsi) & $\langle p_{T}\rangle$ (DY)& 
 $\langle p_{T}\rangle$ (\upsi) &
 $\langle p_{T}^2\rangle$ (DY)  & $\langle p_{T}^2\rangle$ (\upsi)\\
 & (GeV/c) & (GeV/c) & (GeV/c) & (GeV/c) & (GeV/c)$^2$ & (GeV/c)$^2$\\
\hline
 \mbox{p-Be} & 2.77$\pm$0.04 & 3.4$\pm$0.6 & 1.19$\pm$0.03 & 1.5$\pm$0.2 & 1.91$\pm$0.05 &
 2.9$\pm$1.0\\
\hline
 \mbox{p-Al} & 2.75$\pm$0.03 & 3.3$\pm$0.4 & 1.18$\pm$0.01 & 1.4$\pm$0.2 & 1.89$\pm$0.04 &
 2.7$\pm$0.6\\
\hline
 \mbox{p-Cu} & 2.80$\pm$0.02 & 3.0$\pm$0.4 & 1.20$\pm$0.01 & 1.3$\pm$0.2 & 1.96$\pm$0.04 &
 2.3$\pm$0.7\\
\hline
 \mbox{p-Ag} & 2.78$\pm$0.02 & 2.6$\pm$0.3 & 1.19$\pm$0.01 & 1.1$\pm$0.1 & 1.93$\pm$0.03 &
 1.7$\pm$0.4\\ 
\hline
 \mbox{p-W}  & 2.80$\pm$0.03 & 3.0$\pm$0.4 & 1.20$\pm$0.01 & 1.3$\pm$0.2 & 1.96$\pm$0.04 &
 2.3$\pm$0.7\\
\hline
\end{tabular}
\caption{$p_0$ (see Eq.~\ref{eq:pt}), $\langle p_{T}\rangle$ and $\langle p_{T}^2\rangle$ for 
Drell-Yan ($4.5<m_{\mu\mu}<8.0$ GeV/c$^{2}$) and \upsi. The quoted errors for Drell-Yan
are purely statistical, while the values for the \upsi\ include a
systematical error due to the uncertainty in the extrapolation of the Drell-Yan yield into the 
\upsi\ region. 
The total error is anyway dominated by the statistical contribution, due to the
low \upsi\ statistics.}
\label{tab:3}
\end{table}

\newpage

\newpage

\newpage

\newpage

\newpage

\newpage

\newpage

\newpage

\end{document}